\documentclass[12pt,preprint]{aastex}
\usepackage{natbib,graphicx}

\shorttitle{Cosmological Studies with Radio Galaxies and Supernovae}
\shortauthors{Daly et al.}

\begin{document}

\title{Cosmological Studies with Radio Galaxies and Supernovae}
\author{Ruth A. Daly, Matthew P. Mory,\altaffilmark{~}}
\email{rdaly@psu.edu}

\affil{Department of Physics, Penn State University, Berks Campus, P. O. 
Box 7009, Reading, PA 19610}
\author{C. P. O'Dea, P. Kharb, S. Baum,\altaffilmark{~}}
\affil{Rochester Institute of Technology, 54 Lomb Memorial Drive, 
Rochester, NY 14623}
\author{E. J. Guerra\altaffilmark{~}}
\affil{Department of Physics and Astronomy, 
Rowan University, 201 Mullica Hill Rd. Glassboro, NJ 08028}
\and
\author{S. G. Djorgovski\altaffilmark{~}}
\affil{Division of Physics, Mathematics, and Astronomy, 
California Institute of Technology, MS 105-24, 
Pasadena, CA 91125}


\vfill\eject

\begin{abstract}
Physical sizes of extended radio galaxies can be employed as
a cosmological ``standard ruler'', using a previously developed method.
Eleven new radio galaxies are added to our previous sample of nineteen 
sources, forming a sample of thirty objects with redshifts between 
0 and 1.8.  This sample of radio galaxies are used to obtain the best fit
cosmological parameters in a quintessence model in a spatially flat
universe, a cosmological constant model that allows for
non-zero space curvature, 
and a rolling scalar field model in a spatially flat
universe. Results obtained with radio galaxies are
compared with those obtained with different 
supernova samples, and with combined radio galaxy 
and supernova samples.  Results obtained with different
samples are consistent, suggesting
that neither method is seriously affected by systematic errors. 
Best fit radio galaxy and supernovae 
model parameters determined in the different 
cosmological models are nearly identical, and 
are used to determine dimensionless coordinate 
distances to supernovae and radio galaxies, and 
distance moduli to the radio galaxies.  The distance
moduli to the radio galaxies can be combined
with supernovae samples to increase the number of sources,
particularly high-redshift sources, in the samples. 

The constraints obtained here with the combined
radio galaxy plus supernovae data set in the rolling scalar field model
are quite strong.  The best fit parameter values suggest that 
$\Omega_m$ is less than about 0.35, and 
the model parameter $\alpha$ is close to zero; 
that is, a cosmological constant provides a good 
description of the data. 

We also obtain new constraints on the physics of engines that
power the large-scale radio emission.
The equation that describe the predicted size of 
each radio source is controlled by one 
model parameter, $\beta$, which parameterizes
the extraction of energy from the black hole. 
Joint fits of radio galaxy
and supernova samples indicate a best 
fit value of $\beta$ that is very close to a special value for which the
relationship between the braking magnetic field strength and the
properties of the spinning black hole is greatly simplified, and
the braking magnetic field strength depends only upon the spin
angular momentum per unit mass and the gravitational radius
of the black hole.  The best fit value of $\beta$ of 1.5 indicates that
the beam power $L_j$ and the initial spin energy of the 
black hole $E$ are related by $L_j \propto E^2$, and 
that the relationship that might naively be 
expected for an Eddington limited
system, $L_j \propto E$, is quite clearly 
ruled out for the jets in these systems. 

\end{abstract}


\vfill\eject

\section{Introduction}

Recent cosmological studies using CMBR, supernovae, and other types of
astronomical sources and phenomena have greatly improved our understanding
of the recent expansion and acceleration history of the universe.  Whereas
a consistent picture has emerged (the ``concordance cosmology''), in which 
the dynamics of the universe is currently dominated by a mysterious dark 
energy, its physical nature remains one of the key outstanding problems
of physical science today. For a summary of developments in this field
see Ratra \& Vogeley (2007).

Aside from the CMBR, most experimental methods to study the expansion
history of the universe, and thus its matter-energy contents, require samples
of standardisable sources whose distances can be determined in a consistent
way, e.g., the supernovae of type Ia.
It is clear that both low and high redshift sources play an important role
in these studies. Low to moderate redshift sources allow us to define and probe
the acceleration of the universe and the properties of the dark
energy at the current epoch, whereas higher
redshift sources allow us to probe the
properties of the dark energy at earlier epochs and 
possible changes in its properties, which is perhaps
the best path towards understanding its physical nature.
Confidence in luminosity and coordinate distance
determinations, which are the foundation of the studies discussed here, is
bolstered when more than one method yields the same results.  

Like supernovae, powerful radio galaxies are observed out
to redshifts greater than one, and their observable properties 
can be used to determine
the coordinate distances to them, 
or equivalently the luminosity distances or distance moduli (Daly 1994). 
It is thus interesting to study these sources
and to determine whether cosmological results obtained with radio 
galaxies agree with those obtained with supernovae and other methods.  
In addition,
the radio galaxy and supernova samples may be analyzed jointly 
to improve the determinations of the radio galaxy model parameters. 

In this paper, eleven new radio galaxies are combined with a previously studied
sample of nineteen sources, to yield a sample of thirty radio galaxies
suitable for cosmological studies.
This sample is analyzed here in three cosmological models.  
The first model 
allows for quintessence and
non-relativistic matter in a spatially flat universe; 
the second model allows for 
non-relativistic matter, a cosmological constant, and
space curvature; and
the third model allows for a rolling scalar field in a spatially flat
universe (Peebles \& Ratra 1988). All of these 
models are based on the equations of 
General Relativity, and rely upon this as the correct theory 
of gravity. 

The primary objectives are to obtain and compare constraints on model
and cosmological parameters using different samples
of type Ia supernovae, extended radio galaxies, and a combined sample
of supernovae and radio galaxies. Similar results obtained with the 
radio galaxy and supernovae methods, which determine similar quantities
over similar redshifts, will bolster our confidence in each method.
This is because the methods rely upon measurements of entirely different 
quantities that are then applied using completely different
astrophysical arguments, and thus provide independent measures
of coordinate distances and cosmological parameters. 
Results obtained from the combined sample allow strong constraints
to be placed on the radio galaxy model parameter, which provides a direct
link to and diagnostic of the physics of the energy extraction that
produces the large-scale jets in these systems. Finally, if very similar 
best fit
model parameters are obtained in different cosmological scenarios,
tben these parameters can be used to determine the dimensionless
coordinate distance to each source.  The dimensionless coordinate
distances thus obtained are then used for a separate study.  In addition,
the dimensionless coordinate distances to the radio galaxies can be
used to define the distance modulus to each radio galaxy, which can
be combined with those of supernovae to increase the sample sizes. 

The use of powerful extended radio galaxies as a cosmological tool
is reviewed in section 2.  Results obtained with radio galaxies,
supernovae, and combined samples of radio galaxies and supernovae
are presented in section 3.  Determinations of distance moduli to
radio galaxies are presented in section 4.  The main results and
conclusions of the paper are summarized in section 5.   

\section{Radio Galaxies and Cosmology}

Powerful radio sources are a cosmological population which can be observed
out to very high redshifts, reaching beyond the practical limits of supernova
studies.  This makes them in principle a very interesting cosmological probe,
provided that some distance-dependent quantity such as physical size can
be standardized. In section 2, we describe how the maximum size that 
a radio source will reach during its lifetime, $2D_*$, can be determined
from radio observations. 

There are many different types of radio sources such as compact and extended 
radio sources, and radio galaxies and radio loud quasars.  
The use of radio sources for cosmological studies has a long 
and distinguished history including the works of 
Rowan-Robinson (1967), Longair \& Pooley (1969), 
Hoyle \& Burbidge (1970), Kellerman (1972, 1993), Fanaroff \& Longair (1972),
Kapahi (1972, 1985), Rees (1972), 
Readhead \& Longair (1975), Longair (1976),
Wall, Pearson, \& Longair (1980), 
Laing, Riley, \& Longair (1983),
Condon (1984a,b), Gopal-Krishna \& Wiita (1987), 
Pelletier \& Roland (1989), 
Dunlop \& Peacock (1990), Singal (1993), 
Daly (1994), Kayser (1995), Buchalter et al. (1998), 
Guerra \& Daly (1998), 
Kaiser \& Alexander (1999), Gurvits, Kellerman, \& Frey (1999), 
Guerra, Daly, \& Wan (2000), Rawlings (2002), Daly \& Guerra (2002),
Chen \& Ratra (2003), Podariu et al. (2003), 
Jamrozy (2004), Jackson (2004), 
Barai \& Wiita (2006, 2007), and Jackson \& Jammetta (2006). 

Here, very powerful radio galaxies are used as a modified standard 
yardstick for cosmological studies. Radio galaxies rather than
radio loud quasars are selected for study so as to minimize
projection effects; in the standard unified model for radio 
galaxies and radio loud quasars, the sources are intrinsically
the same, but radio galaxies lie close to the plane of the
sky, and radio loud quasars are oriented along the line of sight
to the observer.

The subset of classical double radio galaxies that are very powerful
form a very homogeneous population.  The source properties and structure
are well described by the standard ``twin jet'' model (e.g. 
Blandford and Rees 1974;
Scheuer 1974; Begelman, Blandford, and Rees 1984; Begelman 
\& Cioffi 1989; Daly 1990; Leahy 1991).  
Particles are accelerated and 
material is channeled away from the vicinity of a massive black
hole along relatively narrow, oppositely directed jets, and is deposited
in the radio hot spot.  Here particles are re-accelerated to relativistic
energies and produce synchrotron radiation in the presence of a local
magnetic field. Relativistic plasma flows from the 
radio hot spots, and, as time goes on, the location of the radio hot spot
moves further from the central black hole leaving behind a ``radio bridge''
or ``radio lobe'' of relativistic material. 

\subsection{Overview of the Extended Radio Galaxy (ERG) Method}

The ERG method considered here was proposed by Daly (1994), 
and explored and applied 
by Guerra \& Daly (1998), Guerra, Daly, \& Wan (2000),
Daly \& Guerra (2002), and Podariu et al. (2003).
It is described
in detail in these papers, and is summarized briefly here.

The method is based upon the following observations.
(1) The most powerful extended classical double radio galaxies,
those  with 
178 MHz radio powers greater than about
$(3 \times 10^{26})~h^{-2} \hbox{ W Hz}^{-1} \hbox{ sr}^{-1}$,
have very regular radio bridge 
structure and shape (e.g. Leahy, Muxlow, \& Stephens 1989).
This indicates that the sources are growing at a rate
that is well into the supersonic regime, and that there is
minimal backflow of material within the radio bridge
(Leahy \& Williams 1984; Alexander 1987; Alexander \&
Leahy 1987; Leahy, Muxlow, \& Stephens 1989).  Thus, 
strong shock physics, which is clean and simple,
can be applied to these systems.  It also  
indicates that the velocity 
deduced by studying a source represents the average
rate of growth of the source, that is, corrections for backflow
of material within the bridge are negligible.
(2) The average size of this special class of 
powerful classical double radio galaxies
at a given redshift $\langle D \rangle$ has a small dispersion
independent of the cosmological model,  
as illustrated in Figure 8 of Guerra, Daly, \& Wan (2000).  
This means that $2\langle D \rangle$ provides a measure of the maximum size
that a given source at that redshift will reach during its lifetime. 

The parent population of radio galaxies considered for this
study are from the complete sample of 3CRR radio galaxies with 
178 MHz radio powers greater than about
$(3 \times 10^{26})~h^{-2} \hbox{ W Hz}^{-1} \hbox{ sr}^{-1}$. 
This leads to a sample of 70 radio galaxies (Guerra \& Daly 1998).  
This parent population is used to
define $\langle D \rangle$ in several redshift bins, as described by 
GD98. This measure of the mean or maximum size of a given
source at similar redshift depends upon the cosmological
model through the coordinate distance $(a_or)$, since
the size of a given source is $D \propto (a_or)$.  

Individual sources from this parent population are studied in 
detail to arrive at an independent measure of the average size of an
individual 
source, $D_*$. It turns out that $D_* \propto (a_or)^{4/7-2\beta/3}$,
where $\beta$ is a model parameter to be determined.  The ratio
$R_* \equiv \langle D \rangle/D_*$ is approximately given by 
$R_* \propto (a_or)^{3/7+2\beta/3}$, so requiring that this
ratio remain constant allows a determination of both $\beta$ and
the cosmological parameters that determine $(a_or)$ to a particular
source at redshift z.  When the
fits are run, the full dependence of each factor that enters
into the ratio $R_*$ is, of course, included. The full equation 
for $D_*$ and the way that it was derived is described below
(see also Daly 1990, 1994; GD98; GDW00; DG02).   

Each side of a source grows with an average velocity $v$ for a total
time $t_*$, so a particular 
source will have an average size $D_*$ at the end of it's
lifetime: $D_* \propto v~t_*$.  
(Note that the source velocities are independent of source size,
suggesting that the velocity of a given source is roughly constant
over the lifetime of that source, as described in detail by 
O'Dea et al. 2007.)
The fact that the source sizes have a
small dispersion at a given redshift suggests a cancelation
between that factors that determine $v$ and those that determine
$t_*$. The equations of strong shock physics indicate that 
the overall velocity $v$ with 
which the source lengthens is $v \propto [L_j/(n_a~a_L^2)]^{1/3}$
(e.g. Rawlings \& Saunders 1991; 
Daly 1990), where $n_a$ is the ambient gas density 
and $a_L$ is the radius of the cross sectional area
perpendicular to the direction of growth of the source. 
Thus, the average source size 
depends both of the intrinsic properties, $L_j$ and $t_*$, 
of the AGN that 
powers the outflow,  and those that are extrinsic
to the AGN, $n_a$ and $a_L^2$.  Separating the intrinsic from 
the extrinsic factors we have 
$D_* \propto (L_j^{1/3}t_*)~(n_a a_L^2)^{-1/3}$.  
As shown by Daly (1994), the extrinsic factors
described by the second term have a very weak dependence
on the coordinate distance $(a_or)$ to the source, 
and go approximately as $k_g \propto (n_a~a_L^2)^{-1/3} \propto (a_or)^{-0.1}$.
This parameter can be determined from observations by noting
that $n_a \propto P/v^2$ (e.g. De Young 2002), 
so $k_g \equiv (Pa_L^2/v^2)^{-1/3}$; 
here $P$ is the postshock
pressure in the radio bridge
$P=((4/3)b^{-1.5}+b^2)(B_{min}/24\pi)$ and the 
parameter $b$ allows for offsets of the magnetic field
strength $B$ from minimum energy conditions $B = b~B_{min}$.

Re-arranging terms, we have 
\begin{equation}
D_* \propto vt_* \propto (L_j^{1/3}t_*)k_g~.
\end{equation}
To maintain 
the observed small dispersion
in average source size $\langle D \rangle$, there must be a cancelation between the 
factors intrinsic to the AGN that affect the source size, 
$L_j$ and $t_*$.  Thus, it is proposed that the total lifetime
of the outflow $t_*$ be written as a power-law in the beam
power 
\begin{equation}
t_* \propto L_j^{-\beta/3}~.
\end{equation}
Then, the average size
a given source would have if it were observed over its entire lifetime
is $D_* \propto L_j^{(1-\beta)/3}~(Pa_L^2/v^2)^{-1/3}$,
or $D_* \propto L_j^{(1-\beta)/3}~k_g$.  
Thus, $D_*$ can be determined for an individual source; 
it has one model parameter, $\beta$, and depends upon the cosmological
model through its depedence on $(a_or)$.  

This parameterization of the total time 
the AGN produces jets $t_* \propto L_j^{-\beta/3}$   
includes as a special case the 
relationship expected if the outflow
is Eddington limited.  For an Eddington limited 
system, the lifetime is independent of the beam
power, so a value of $\beta = 0$
is expected, and
we will be able to test whether the outflows are Eddington limited. 

The implications of the value of $\beta$ for physics quite
close to the AGN are described by Daly \& Guerra (2002), 
and are discussed in more detail here in section 3.1. 
The relationship $t_* \propto L_j^{-\beta/3}$ 
fits quite nicely in standard magnetic braking models
to power the outflows from AGN (e.g. Blandford 1990; see
section 3.1).   

To write $D_*$ in terms of empirically determined quantities
$a_L$, $v$, and $P$, we note that 
$D_* \propto v~t_* \propto v~L_j^{-\beta/3}$ and 
$L_j \propto (v/k_g)^3$, so 
\begin{equation}
D_* \propto k_g^{\beta}~v^{1-\beta}
\propto (a_L^2P)^{-\beta/3}~v^{1-\beta/3}.
\end{equation}
The dependence of $D_*$ on cosmological parameters enters
through the dependence of each of the empirically determined
quantities, $a_L$, $P$, and $v$, on the coordinate distance, and 
is approximately given by $D_* \propto (a_or)^{4/7-2\beta/3}$ Daly (1994).
Of course, when the full computation requiring that the 
ratio $\langle D \rangle/D_*$ remain constant is carried out, 
the full dependence of each quantity $a_L$, $P$, and $v$ on
the coordinate distance is included.

\section{Results}

Here,
we consider three standard models: 
a quintessence model in a spatially flat universe; 
a lambda model that allows for non-zero
space curvature, non-relativistic matter, and a cosmological constant; 
and a 
rolling scalar field model in a spatially flat universe.  
The supernovae samples considered here are
the 192 supernovae from Davis et al. (2007), the 182 supernovae sample
of Riess et al. (2007), and the 115 supernovae sample of Astier et al.
(2006). There is significant overlap between these samples; 
for example, the Davis et al. (2007) sample includes the Essence data of
Wood-Vasey (2007) and the high redshift data of Riess et al. (2007).
Different samples are studied to be able to compare results obtained
with each.  In addition, the model parameters determined for each
sample can be applied to that sample to solve for the dimensionless
coordinate distances to the supernovae in that sample.  
The 30 radio galaxies studied here include the eleven new
radio galaxies presented by Kharb et al. (2008), with details on 
individual source properties given by O'Dea et al. (2007), and 
the nineteen radio galaxies previously studied by Guerra, Daly,
\& Wan (2000) and Guerra \& Daly (2002) with the source 3C427.1
removed, as discussed by Podariu et al. (2003).

The best fit parameters obtained in these models are listed
shown in Tables 1, 2, and 3.   
The fits are done in the standard way for radio galaxies
and supernovae separately.  
For the radio galaxies, we use the relationship  
\begin{equation}
R_* = \langle D \rangle/D_* = 
k_0y^{(6\beta-1)/7}(k_1y^{-4/7}+k_2)^{\beta/3-1}~,
\end{equation}
where the cosmological model enters through the determination
of 
the dimensionless coordinate distance $y$ to each source,
which is simply related to 
the coordinate distance $(a_or)$, $y=(H_0/c)(a_or)$, and
we have the standard equations
$y=\int dz/E(z)$, $E(z) = H(z)/H_0$, $H(z) = (1/a)(da/dt)$
is the expansion rate of the universe at a given redshift, 
and $k_0$, $k_1$, and $k_2$ are observed quantities
(see Daly \& Djorgovski 2003 and the appendix of Guerra \& Daly 2002). 
We minimize 
the difference between $ ln(\langle D \rangle/D_*)$ and 
a constant, $\kappa_{RG}$, 
as described in detail by Guerra \& Daly (1998), 
to obtain the best fit values of $\beta$,
$\kappa_{RG}$, and cosmological parameters.  
The ratio $\langle D \rangle/D_*$ obtained for the best fit values of
cosmological parameters and $\beta$, and normalized to unity using
the best fit value of $\kappa_{RG}$ is shown in Figure 1
for the quintessence model. As in Daly \& Guerra (2002), 
an offset from minimum energy conditions
of $b=0.25$ has been adopted, and the results are insensitive
to the value of $b$ assumed; similar results are obtained for
$b=1$.  
For the supernovae, we use the relationship
\begin{equation}
\mu = \kappa_{SN} + 5log10[y(1+z)],
\end{equation}
where $\kappa_{SN}$ is a constant to be fitted for each
of the supernvae samples, and,
as above, $y$ is the dimensionless coordinate distance,
$y=(H_0/c)(a_or)$.  We
minimize the difference between the observed and predicted
values of $\mu$ to obtain the best fit values for cosmological 
parameters and the constant $\kappa_{SN}$.  
The parameter $\kappa_{SN}$ can be related to
the effective Hubble constant adopted by the supernovae
group to obtain their values of $\mu$,
$\kappa_{SN}=25-5log10(H_0/c)$, where $H_0$ is in units
of km/s/Mpc, and c is in units of km/s.  For example,
a value of $H_0 = 70 \hbox{ km s}^{-1} \hbox{ Mpc}^{-1}$ 
translates to a value of $\kappa_{SN}$ of about 43.15. 
This approach is convenient since effective values of 
$H_0$ adopted and used in supernovae studies
are often not included in publications and 
are often otherwise unavailable for the supernovae
samples; the approach used here by-passes the need to know the
specific value of $H_0$ adopted and applied to a particular supernovae
sample.  The exception is the supernovae sample of 
Astier et al. (2006) who adopted a value of 
$H_0 = 70 \hbox{ km s}^{-1} \hbox{ Mpc}^{-1}$, 
corresponding to a value of $\kappa_{SN}=43.15$;
the values we recover here (see Tables 1, 2, and 3)
are consistent with the input value of Astier et al. (2006).    
Once the best fit value of $\kappa_{SN}$ has been determined for a 
particular data set, this value and the observed value
of $\mu$ can be substituted into equation (5) to solve for the
dimensionless coordinate distance to the sources in that sample; the values
thus obtained are listed in Daly et al. (2008). 
Similarly, once the best fit values of the constants
$\kappa_{RG}$ and $\beta$ have been obtained, equation (4) can
be used to solve for the dimensionless coordinate distance y to 
each radio galaxy.  Then, equation (5) can be used to define an
effective distance modulus $\mu$ to each radio galaxy so that
the radio galaxies can be added to and analyzed with the 
supernovae samples.  Of course, in doing this, the best fit
value of $\kappa_{SN}$ obtained for a particular supernovae sample
is used to obtain values of $\mu$ for the radio galaxies that are
to be added to that sample.

To obtain the best fit parameters
for the joint fits, the chi-square for each set of cosmological parameter
and model parameter values is obtained by adding the chi-square obtained 
for the radio galaxies and
supernovae, and finding the minimum chi-square of the joint fit.
That is, we require that a single set of cosmological parameters
and model parameters $\beta$, $\kappa_{RG}$, and $\kappa_{SN}$ 
describe both the
radio galaxy and supernovae data sets simultaneously, and
find the best fit values of parameters that minimizes this
total chi-square.  This is quite helpful in constraining the 
model parameter $\beta$, as discussed below. 

Each of 
the radio galaxy fits involve four parameters,
two cosmological parameters and two model parameters 
($\kappa_{RG}$ and $\beta$),
or 26 degrees of freedom for 30 radio galaxies.  
The supernovae fits involve three parameters,
two cosmological parameters and one model parameter ($\kappa_{SN}$), since
the supernovae model parameter that relates the rate of decline
of the light curve to the peak supernovae brightness is obtained separately. 
The joint supernovae
and radio galaxy fits involve five parameters, two cosmological
parameters and three 
model parameters ($\kappa_{SN}$, $\kappa_{RG}$, and $\beta$).
The reduced $\chi^2$ for the best fit parameters 
is of order unity for radio galaxy sample alone, for most of the 
supernovae samples alone, and for the joint supernovae and radio
galaxy fits, indicating
that the model provides a good description of the data for the 
best fit parameters listed, though the reduced $\chi^2$ obtained for the 
182 supernovae is a bit low.  Values of cosmological 
parameters obtained here are consistent with those obtained
and published by each of the supernovae groups. In addition
to the best fit values of cosmological parameters, we also list the
best fit values of the model parameters, 
$\kappa_{SN}$, $\kappa_{RG}$, and $\beta$, 
which are needed to solve for the dimensionless coordinate distance
to each source. It is also interesting to compare best fit parameter
values obtained with radio galaxies alone, supernovae alone,
and the combined radio galaxy and supernovae samples. 

\begin{figure}
\plotone{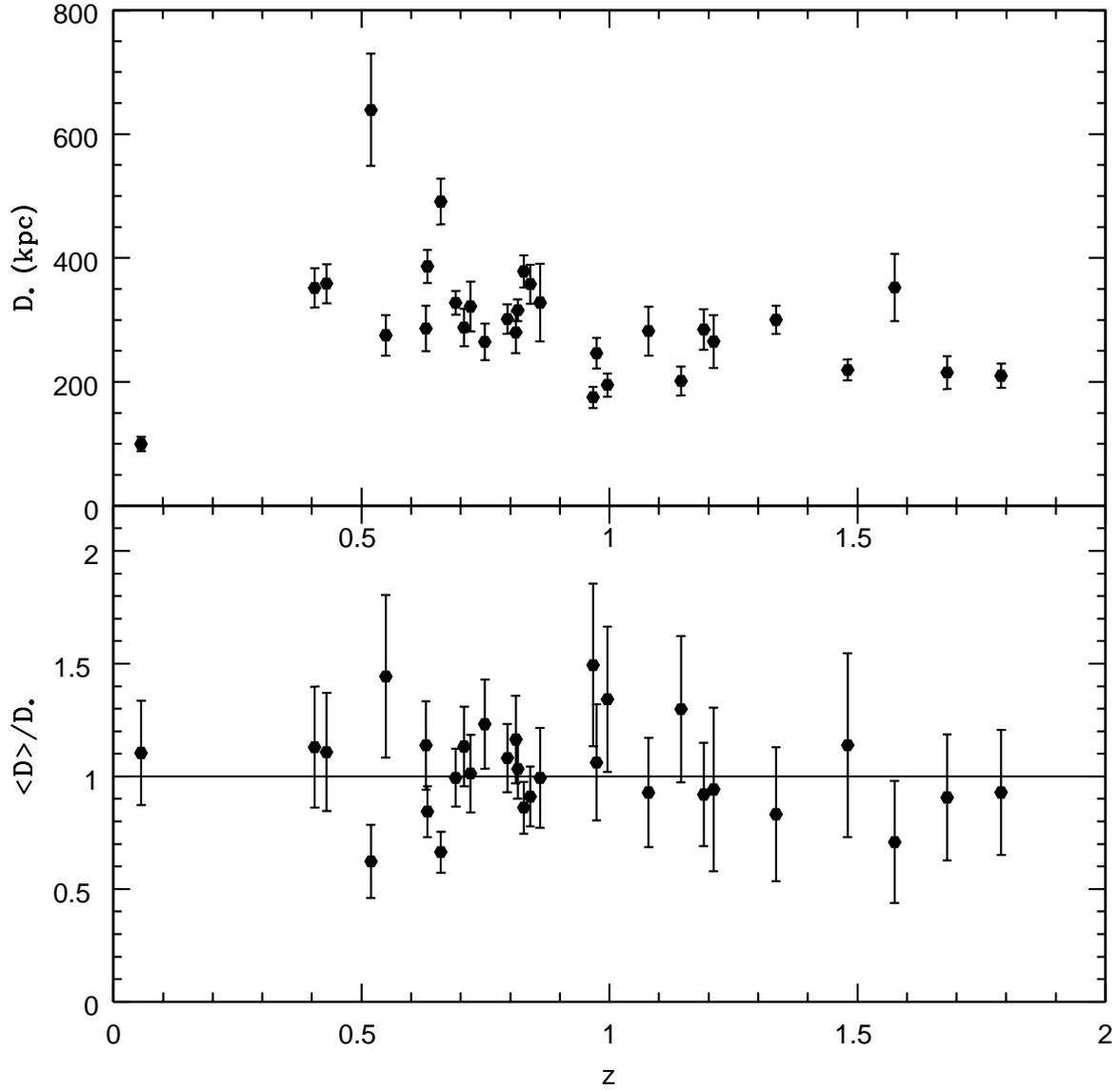}
\caption{Values of $D_*$ and the ratio
$\langle D \rangle/D_*$ for the 30 radio galaxies obtained using the  
the best fit parameters of the quintessence model for
radio galaxies alone and normalized using the best fit value
of $\kappa_{RG}$. } 
\label{dstar}
\end{figure}

Values of $D_*$ and the ratio $\langle D \rangle/D_*$ are shown 
in Fig. \ref{dstar} for the best
fit parameters obtained with the 30 radio galaxies alone in the
quintessence model (see Table 2), normalized using the 
best fit value of $\kappa_{RG}$ to have a value of unity. 
The values of $D_*$
clearly change with redshift as expected; they should
change with redshift in the same way as $\langle D \rangle$, which is
shown in Fig. 8 of Guerra, Daly, \& Wan (2000).  The ratio
$\langle D \rangle/D_*$ is independent of redshift, as predicted in the model. 

\begin{figure}
\plotone{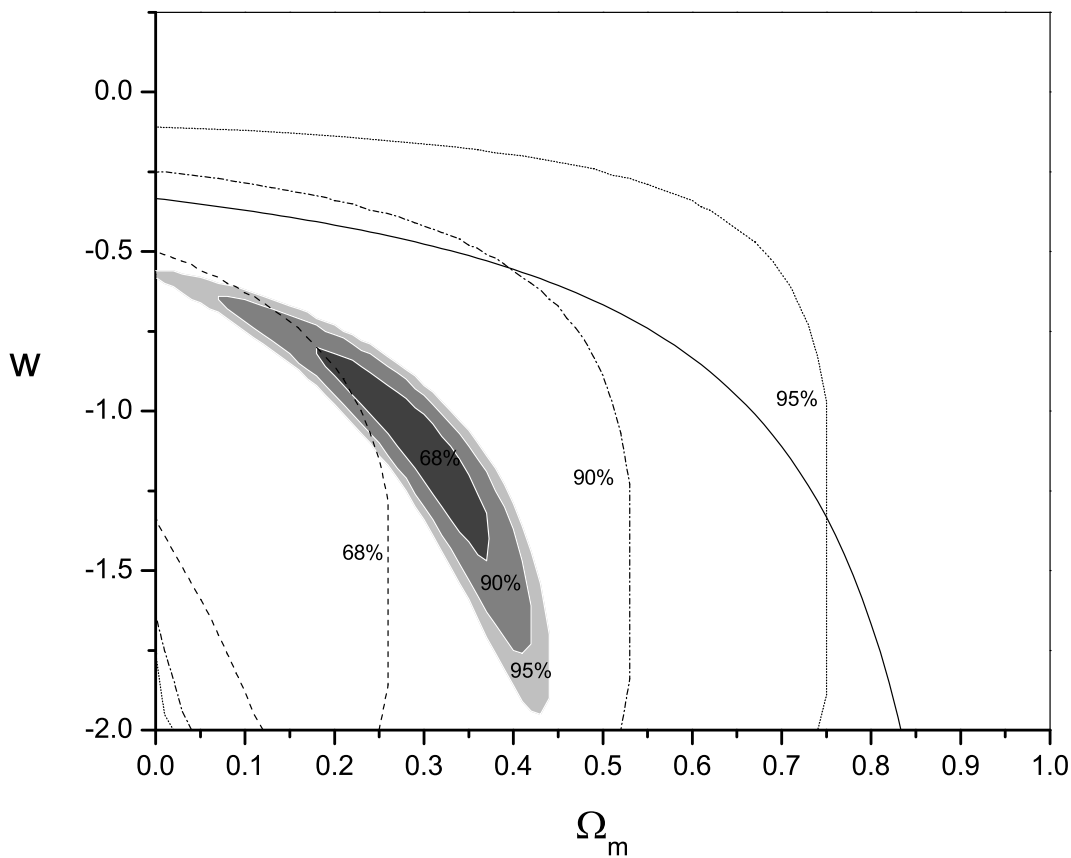}
\caption{Constraints on the equation of state parameter w and $\Omega_m$
obtained in a spatially flat 
quintessence model with 30 radio galaxies alone (dashed lines)
overlayed 
with those obtained with 
the combined sample of 192 supernovae and 30 radio galaxies 
(solid contours). The solid line 
separates an accelerating from a decelerating universe; points below 
the line are parameter values for which the universe is accelerating
at the current epoch in this model. } 
\label{Q30rgommw}
\end{figure}

\begin{figure}
\plotone{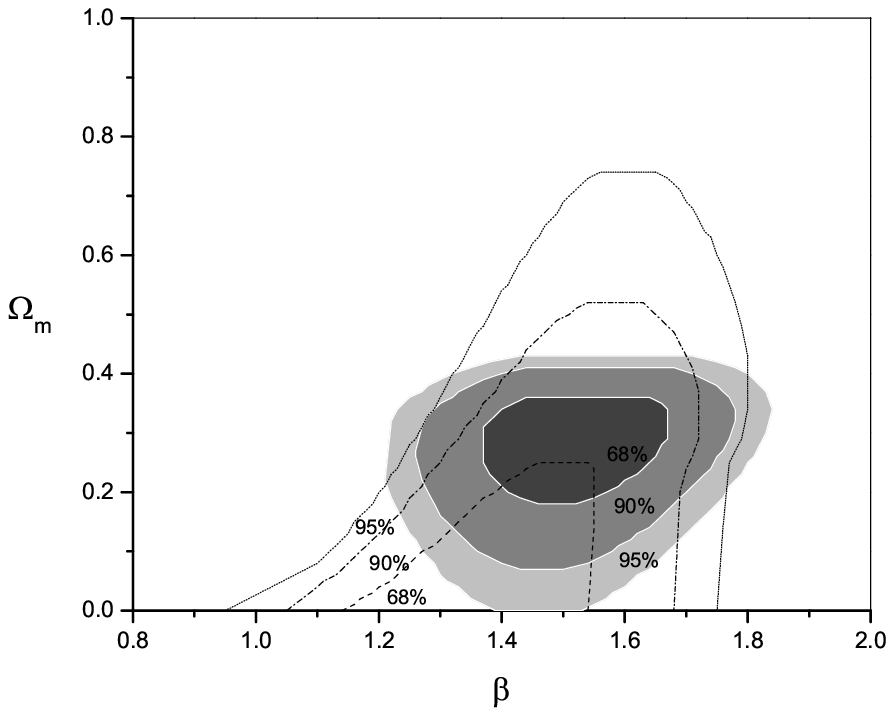}
\caption{As in Figure \ref{Q30rgommw} for the parameters 
$\beta$ and $\Omega_m$.  } 
\label{Q192snp30rgbomm}
\end{figure}

\begin{figure}
\plotone{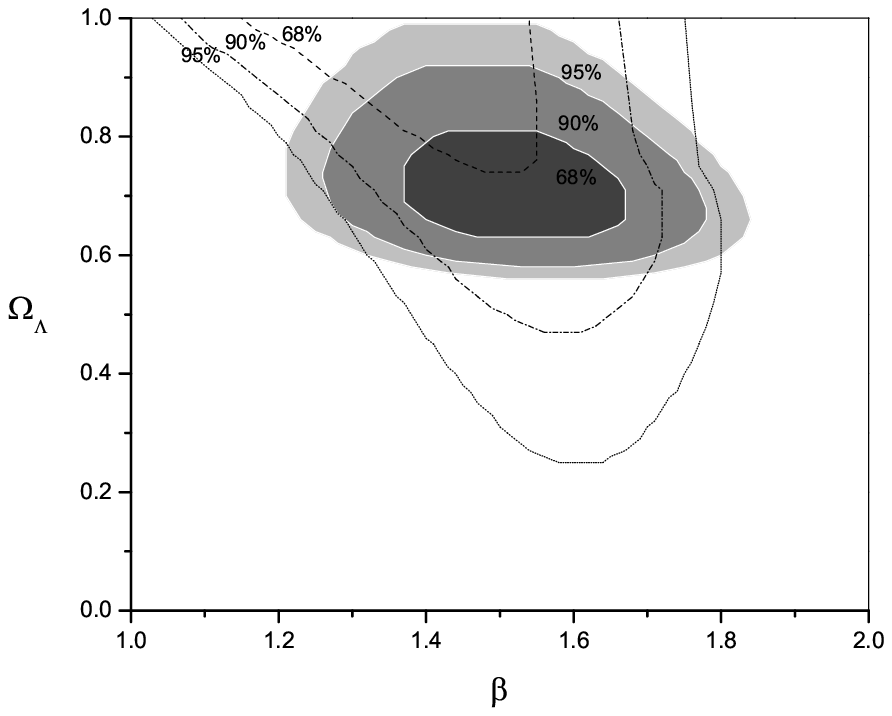}
\caption{As in Figure \ref{Q30rgommw} for the parameters 
$\beta$ and $\Omega_{\Lambda}$.} 
\label{Q192snp30rgbomL}
\end{figure}

\begin{figure}
\plotone{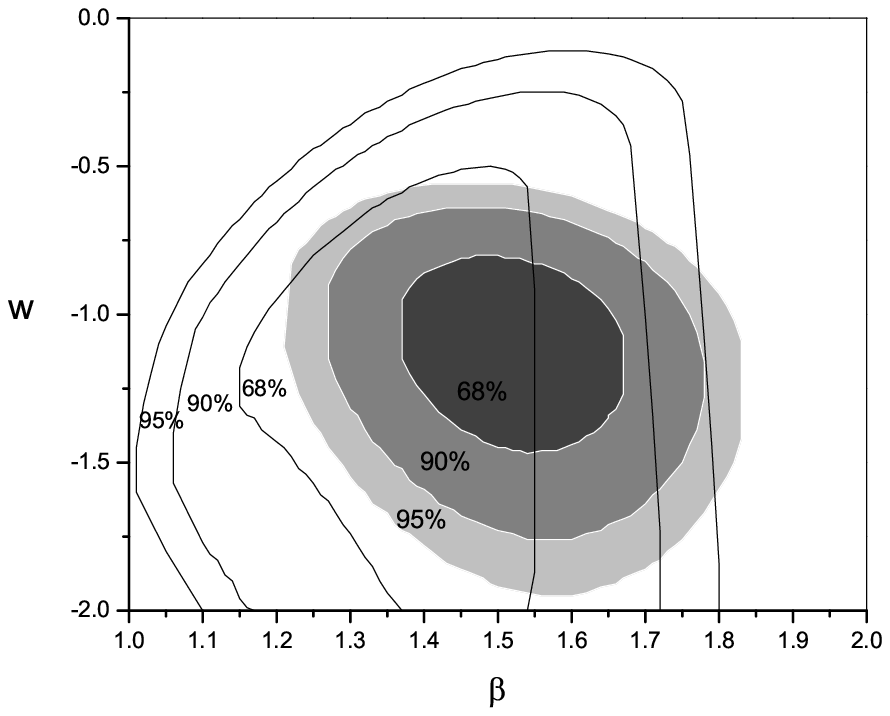}
\caption{As in Figure \ref{Q30rgommw} for the parameters 
$\beta$ and $w$.} 
\label{Q192snp30rgbw}
\end{figure}

Results obtained in a quintessence model assuming a spatially flat
universe are shown in Figures \ref{Q30rgommw} to \ref{Q192snp30rgbw}.
Results obtained with radio galaxies alone are overlayed on those
obtained with the combined sample of supernovae and radio 
galaxies, which are quite similar to those obtained with 
supernovae alone.  The radio galaxy results 
are consistent with but weaker than those
obtained with supernovae alone; radio galaxies alone
indicate that the universe is accelerating today 
with about 90\% confidence. 
The fact that consistent
results are obtained with two completely independent methods
applied to sources with similar redshifts suggests that
systematic errors are not a major problem for either method.
The value of $w$ obtained with radio galaxies alone is slightly larger
than $-1$, the best fit values for the supernovae samples are slightly 
less than $-1$, and the best fit values for the combined samples
are slightly larger than $-1$; apparently the radio galaxies tend to pull the 
supernovae samples to slightly larger values of $w$.  
Again, the radio galaxy parameter $\beta$  
has no covariance with cosmological parameters.

\begin{figure}
\plotone{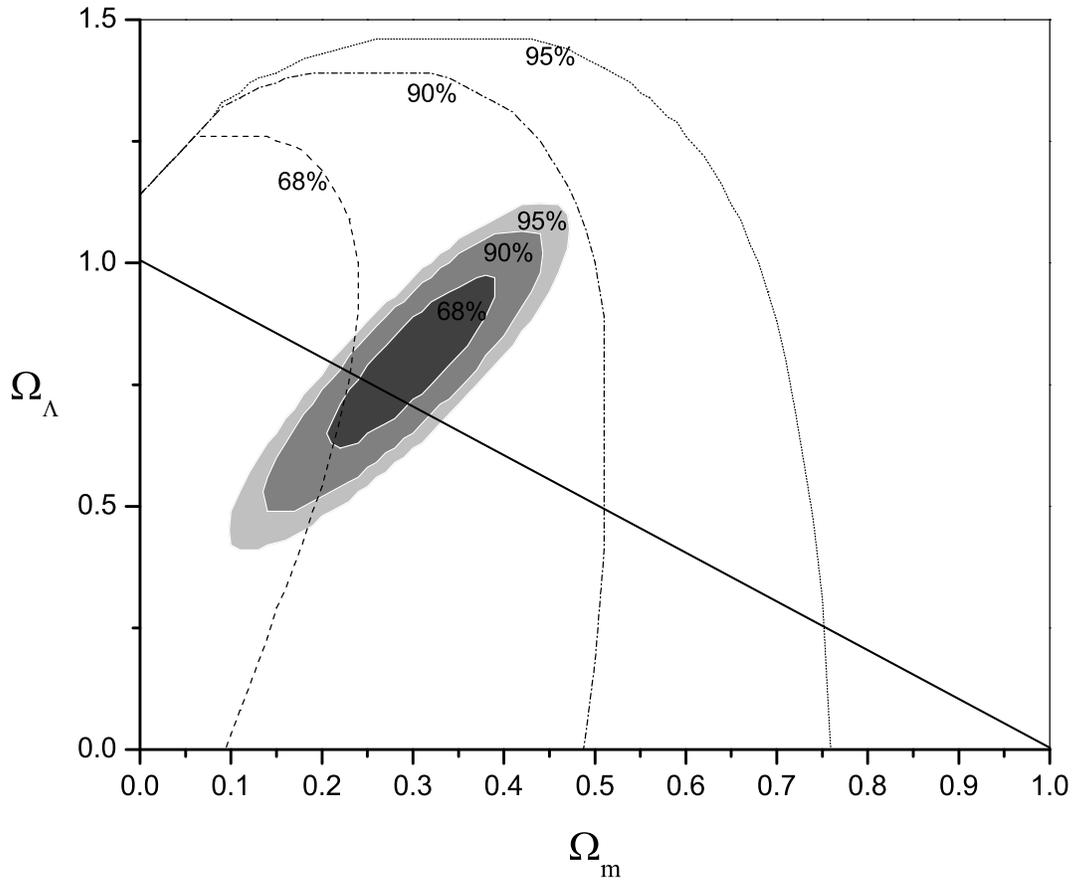}
\caption{Constraints obtained with 30 radio galaxies alone 
(indicated by dashed lines) overlayed 
with those obtained with 
the combined sample of 192 supernovae and 30 radio galaxies 
(indicated by solid contours) 
in a model 
that allows for space curvature, non-relativistic matter, and a cosmological
constant.} 
\label{KrgsnPrgommomL}
\end{figure}

\begin{figure}
\plotone{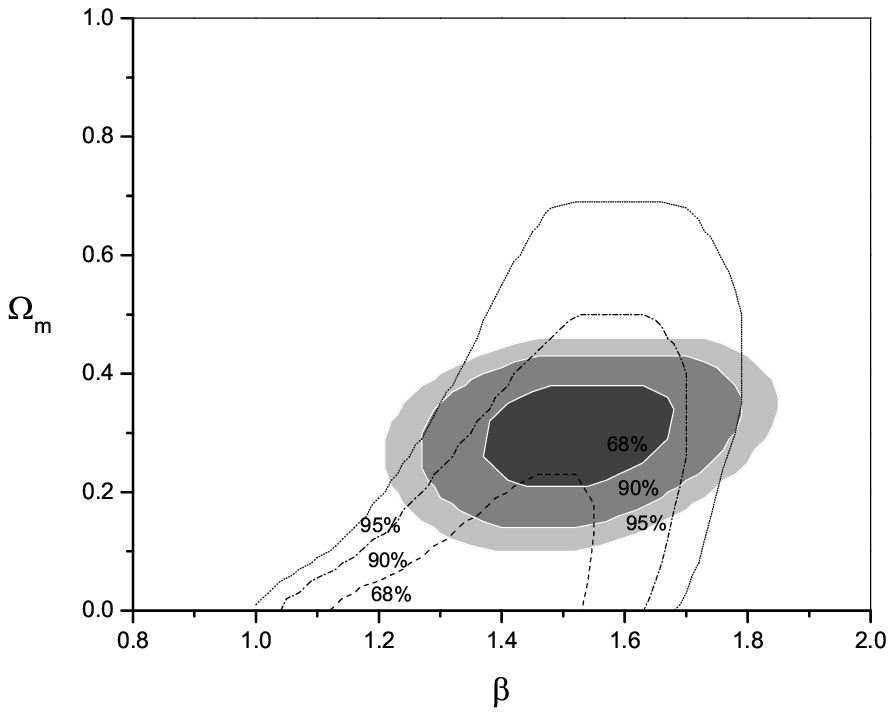}
\caption{As in Figure \ref{KrgsnPrgommomL} for the parameters
$\beta$ and $\Omega_m$.} 
\label{KrgsnPrgbomm}
\end{figure}

\begin{figure}
\plotone{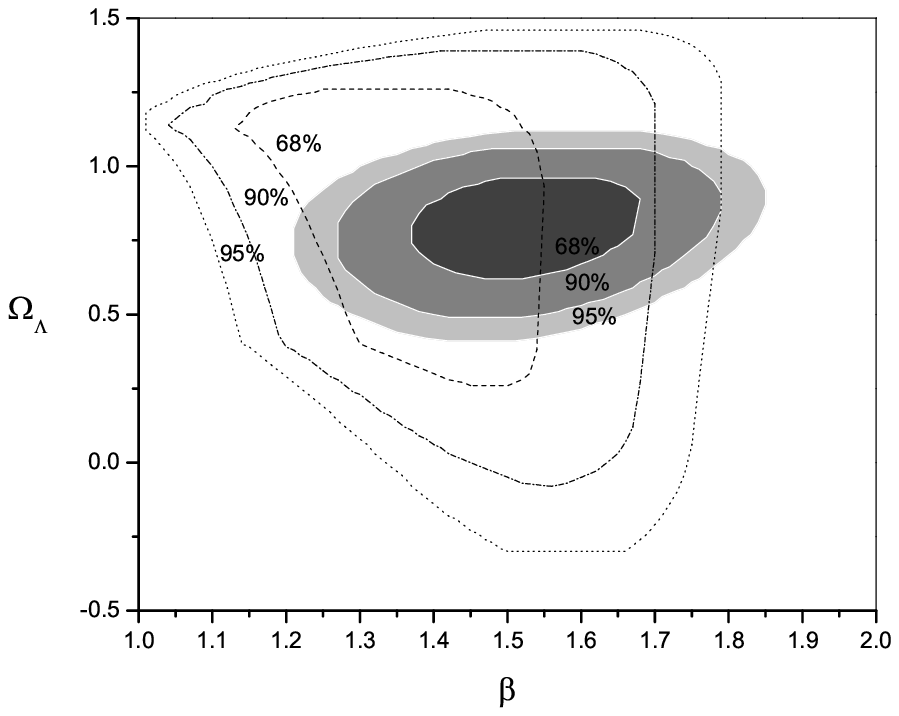}
\caption{As in Figure \ref{KrgsnPrgommomL} for the parameters $\beta$ and
$\Omega_{\Lambda}$.} 
\label{K192snp30rgbomL}
\end{figure}

\begin{figure}
\plotone{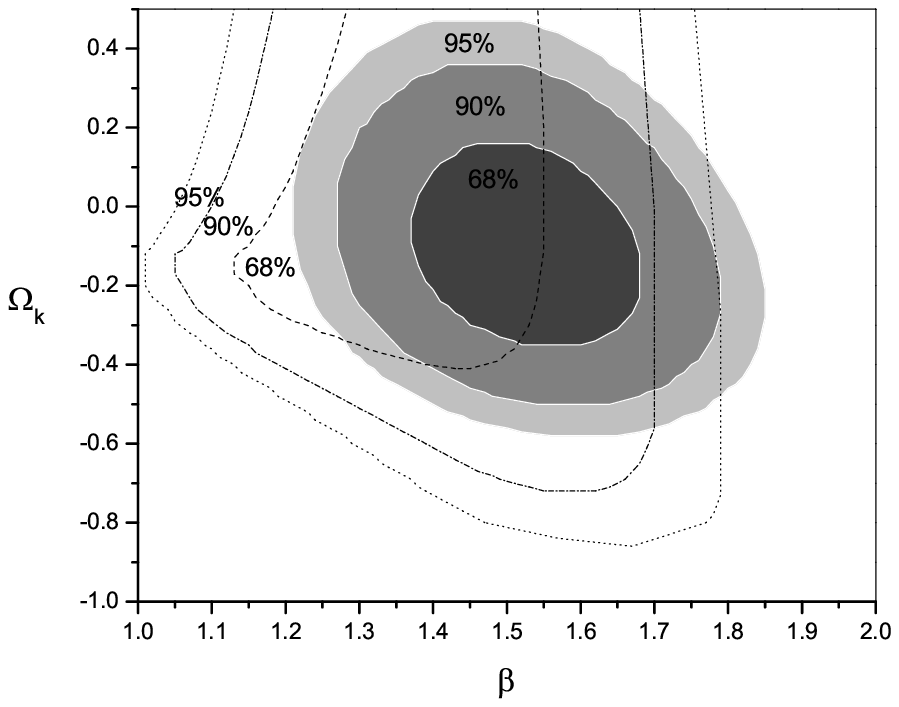}
\caption{As in Figure \ref{KrgsnPrgommomL} for 
the parameters 
$\beta$ and $\Omega_{k}$.}
\label{K192snp30rgbomk}
\end{figure}

Results obtained in a lambda model with space curvature are shown
in Figures \ref{KrgsnPrgommomL} to \ref{K192snp30rgbomk}. Radio galaxies
alone constrain $\Omega_m$ to be less than about 0.5 at 
90\% confidence (see Figure \ref{KrgsnPrgommomL}).  
The combined radio galaxy and Davis et al. (2007)
supernovae sample provide interesting constraints in the 
$\beta$ as shown in Figures 
\ref{KrgsnPrgbomm}, \ref{K192snp30rgbomL}, and 
\ref{K192snp30rgbomk}. Negative $\Omega_k$ indicates
positive space curvature, and the results are 
consistent with zero space curvature. 
As stressed by Tegmark et al. (2006), Wright (2006), and
Wang \& Mukherjee (2007), 
it is only by combining
constraints obtained with different methods that tight constraints
can be placed on space curvature; as seen here, supernovae alone
and radio galaxies plus supernovae, do not place tight constraints
on space curvature. Constraints on
the radio galaxy model parameter $\beta$ obtained with with radio 
galaxies alone are overlayed on these figures,
and are consistent with, though weaker than, those obtained with 
the full sample or with supernovae alone, 
suggesting that neither method is plagued by
systematic errors at this level of accuracy. 
There is no covariance between the model parameter $\beta$ and
cosmological parameters, confirming the results of Daly \&
Guerra (2002). The implications of these constraints on $\beta$ 
are discussed in Section 2.2.

\begin{figure}
\plotone{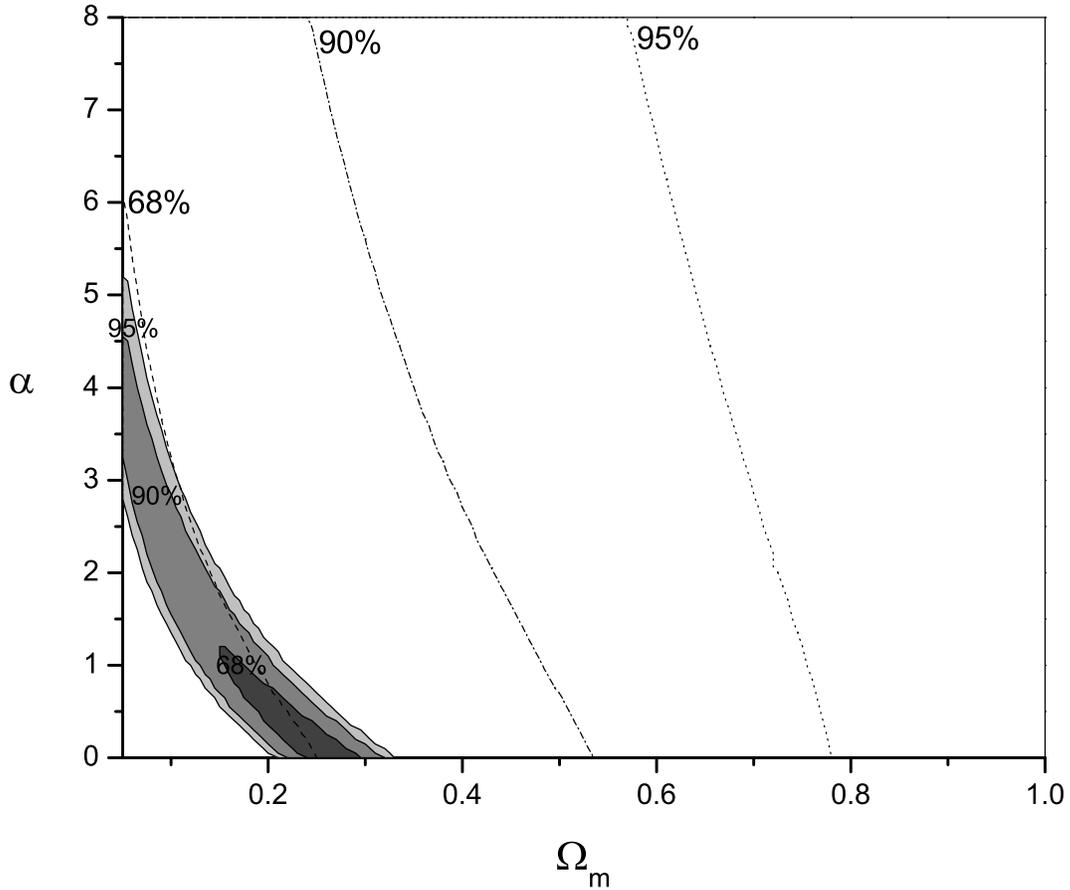}
\caption{Constraints 
obtained in a spatially flat rolling scalar field model  
with 30 radio galaxies alone (dashed lines)
overlayed 
with those obtained with 
the combined sample of 192 supernovae and 30 radio galaxies 
(solid contours) on the 
model parameter $\alpha$ and
$\Omega_m$.} 
\label{A30rgp192SNommalpha}
\end{figure}

Results obtained in the rolling scalar field model of Peebles \&
Ratra (1988) 
are shown in Figure \ref{A30rgp192SNommalpha}. These results
are significantly tighter than those reported by Samushia, Chen, \& 
Ratra (2007).   
The results obtained with the 
joint sample are strong, and suggest that $\alpha$ is close to zero,
that is, a cosmological constant provides a good description of
the data.  

\subsection{Implications of $\beta = 1.5$}

Constraints on the radio galaxy model parameter $\beta$ indicate
that there is little covariance between $\beta$ and  
cosmological parameters, and 
the value of $\beta$ is insensitive to the
cosmological model.  
By combining the radio galaxy and supernovae data, $\beta$ can be 
more tightly constrained.  It is very interesting indeed that 
for the joint fits $\beta = 1.5 \pm 0.15$.  This is a very special
value of $\beta$, as discussed by Daly \& Guerra (2002), and
may explain why these sources provide such an accurate cosmology tool.

The equations of Blandford (1990), $L_j \propto
(a/m)^2~B^2~M^2$ for the beam power and 
$E_* \propto (a/m)^2 M$, where $B$ is the magnetic field strength, 
$a$ is the spin angular momentum per unit mass, m is the 
gravitational radius, and M is the mass of the black hole
that powers the outflow with beam power $L_j$ and total energy $E_*$.
These are combined with 
the empirically derived relationship $E_* = L_jt_* \propto L_j^{1-\beta/3}$
suggested by Daly (1994) to obtain the magnetic field strength
$B \propto M^{(2\beta-3)/2(3-\beta)}~(a/m)^{\beta/(3-\beta)}$
(Daly \& Guerra 2002).   When $\beta = 1.5$, this simplifies to
$B \propto (a/m)$, so the magnetic field strength depends only upon
the spin angular momentum per unit mass and the gravitational radius
of the black hole, and is independent of the black hole mass $M$.  
The fact that our empirically determined value
of $\beta$ is very close to this special value is intriguing,
and may suggest that some physical process is driving the 
magnetic field strength to some maximum or limiting value set by 
$(a/m)$. When the field strength reaches this value, the energy is released
in the form of the jets with beam power $L_j$ that remains roughly constant
over the outflow lifetime $t_*$ releasing a total energy $E_*$ through
the directed jets. 

A value of $\beta$ of 1.5 not only implies that $B \propto (a/m)$;
this, in turn, implies that, $L_j \propto E_*^2$,
and $t_* \propto L_j^{-1/2}$, or $t_* \propto E_*^{-1}$.
The last three relationships follow directly from the
definition of $\beta$, $t_* \propto L_j^{-\beta/3}$
and the relationship $E_* = L_jt_*$.  A value of $\beta$ of 1.5
also implies that 
$L_j \propto (a/m)^4 M^2$ or $L_j \propto B^4M^2$.     
The relationship naively expected for an Eddington limited
system is $L_j \propto M$, so $t_* \propto E_*/M$, which 
only depends upon an efficiency factor when $E_* \propto M$.
Thus, for an Eddington limited system, $t_*$ is 
does not depend explicitly on $L_j$, and would require a value
of $\beta$ of zero, which is quite clearly ruled out
for these systems.   
Given that the most likely source of energy for these
systems is spin energy of a rapidly rotating black hole,
which could originate from the orbital energy of
two black holes that merge, perhaps it is not surprising that
the outflows have little to do with the Eddington luminosity
(e.g. Blandford 1990). 

Studies have shown a close relationship between the radio luminosity 
(jet power) and optical emission line luminosity (AGN ionizing luminosity) 
(Baum \& Heckman 1989; Rawlings \& Saunders 1991; Xu, Livio \& Baum 1999;
Willott et al 1999). 
This may suggest a relationship between the jet lifetime of the
source and the lifetime of the optically bright AGN.  It is
possible that 
both the jet lifetime and the lifetime of the optically bright AGN
are not Eddington-limited, but have a power-law relationship
to the total energy and beam power.

\section{Distances and Distance Moduli to Radio Galaxies}

There are several ways to obtain the dimensionless coordinate
distance $y$ to each source.  For supernovae, one way to is apply 
the best fit value of the constant 
$\kappa_{SN}$ obtained for the sample,
and apply it to the relation 
$y=(1+z)^{-1}~10^{(\mu - \kappa_{SN})/5}$, with the
uncertainty of $y$ given by 
$\sigma_y = [\sigma_{\mu}~y ~ln(10)]/5$, where $\sigma_{\mu}$ is the
uncertainty in the distance modulus $\mu$ to the source, which 
follows from equation (5).   
Similarly, 
one way to obtain the dimensionless coordinate distance $y$ 
to each radio galaxy 
is by applying the best fit
values of $\kappa_{RG} = ln (R_*)$ and $\beta$, and solving 
for the value of y for which $\langle D \rangle/D_* = R_*$ for that source
using equation (4).   
The values of y listed in
Table 4 for 30 radio galaxies are obtained using the 
best fit values of $\kappa_{RG}$ and $\beta$ from  
the joint fits of 192 supernovae and 30 radio galaxies, where
the average of the best fit values 
from Tables 1, 2, and 
3 for the relevant samples 
were used. These dimensionless coordinate distances 
are then converted to distance moduli and their uncertainties 
using the best fit value of $\kappa_{SN}$ for 
the Davis et al. (2007) sample
of 192 supernovae (labeled $\mu_D$ and $\sigma_{\mu(D)}$)
using equation (5).  
Values of $y$ were also obtained (but are not listed) using the 
the best fit values of $\kappa_{RG}$ and $\beta$ for the joint
182 supernovae and 30 radio galaxy fits and converted to a
distance modulus and its uncertainty using the best fit value
of $\kappa_{SN}$ for the Riess et al. (2007)
sample; these are listed in Table 4 
and are labeled $\mu_R$ and $\sigma_{\mu(R)}$.
The distance moduli $\mu_D$ for the radio 
galaxies can be combined with those for supernovae lisetd by 
Davis et al. (2007), 
and the distance moduli $\mu_R$ for the radio galaxies can be
combined with those for the supernovae listed by Riess et al. (2007).  

The values of the constants that are used to obtain the dimensionless
coordinate distances are rather insensitive to the model used to obtain
them; similar results are obtained in the lambda model with space space
curvature, the quintessence model, and the rolling scalar field
model, and average values of the constants obtained in the 
context of these cosmological models 
relevant to each sample was used
to obtain the values listed in Table 4. 
There are two alternate
ways to obtain the dimensionless coordinate distances to the supernovae
and radio galaxies that do not require the use of the best fit
model parameters $\kappa_{SN}$, $\kappa_{RG}$, and $\beta$.  These methods
yield results very similar to those obtained using the best fit parameters.  
The alternate methods start with the values of $\mu$ for supernovae
and $\langle D \rangle/D_*$ for radio galaxies, obtain the 
luminosity distance or the coordinate distance $(a_0r)$ to each source,
which is in units of Mpc, and then solve for the dimensionless
coordinate distance using the equation $y = (H_0/c)(a_0r)$ by 
finding a way to determine $H_0$ that is appropriate for that sample,
which often is 
not stated in the data papers. This value of $H_0$ that was input 
needs to be removed, and can be obtained by
fitting the low redshift data to the Hubble law, or by requiring that
at zero redshift the function $E(z) = H(z)/H_0$, 
described by Daly \& Djorgovski
(2003)
(see also Daly et al. 2008), is equal to one.  It turns out that the 
values of $y$ obtained with any of these three methods are very similar. 

\begin{figure}
\plotone{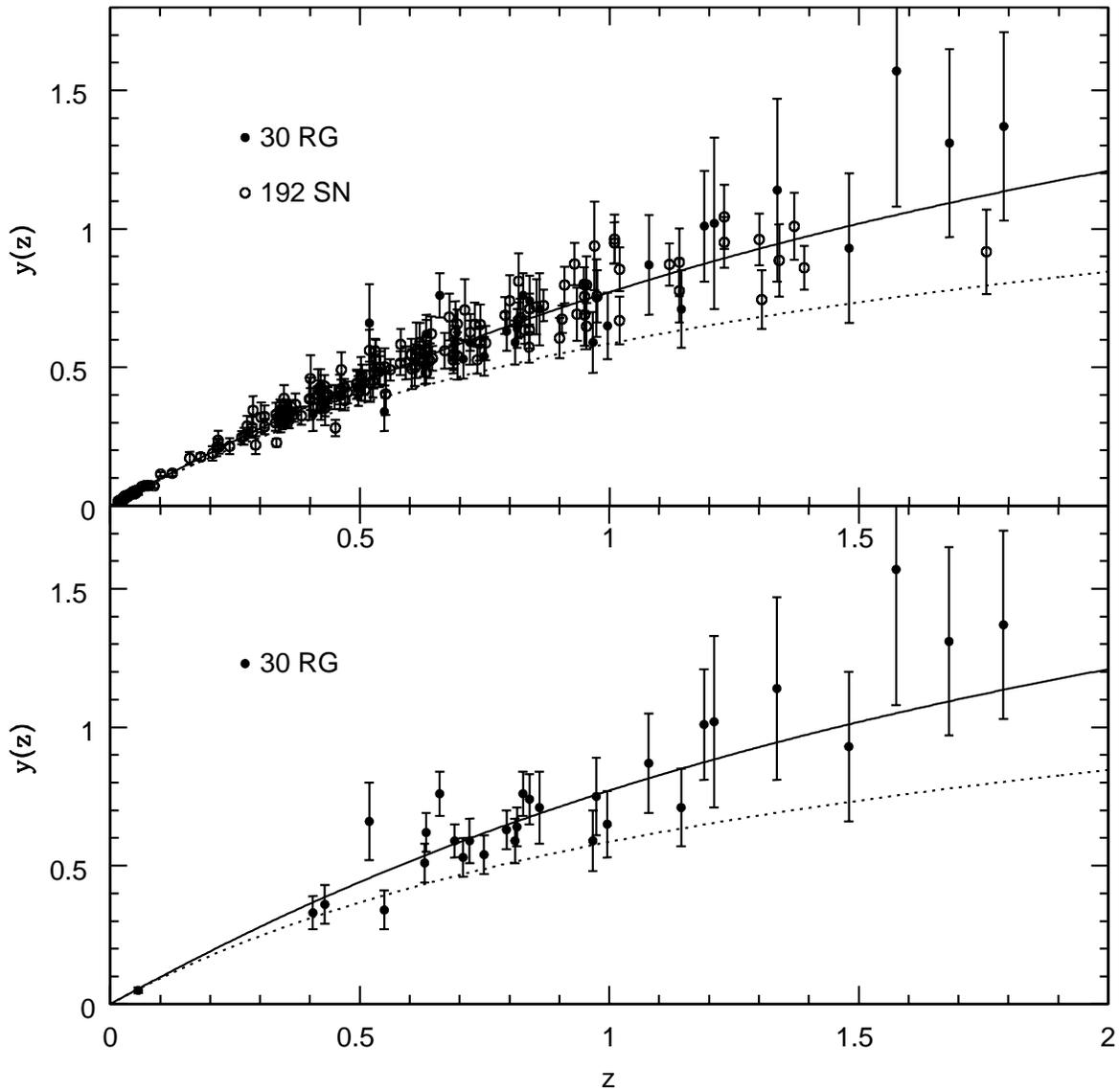}
\caption{Dimensionless coordinate distances to 30 radio galaxies compared with
those to the 192 supernovae of Davis et al. (2007). The dashed curve indicates
y(z) expected in a flat matter dominated universe with $\Omega_m=1$, and the 
solid curve indicates that expected in a flat lambda dominated universe
with $\Omega_m = 0.3$ and $\Omega_{\Lambda}=0.7$; note that $y(z)$ is
independent of $H_0$. Clearly, the data are 
well described by a cosmological constant with $\Omega_{\Lambda}=0.7$. } 
\label{yofz}
\end{figure}

The dimensionless coordinate distances to the 30 radio galaxies
and 192 supernovae of Davis et al. (2007) are shown in
Figure \ref{yofz}; The supernovae data set includes the essence
supernovae presented by Wood-Vasey (2007), the legacy supernovae 
presented by Astier et al. (2006), and 
the high redshift HST supernovae presented by Riess et al. (2007).

Although the radio galaxy and supernovae methods are very different, 
there is good agreement
between the dimensionless coordinate distances to sources at
similar redshift.  The supernovae distances rely on optical
observations for rather short lived events, whereas the
the radio galaxy distances rely on radio observations for
sources that have lifetimes on the order of millions of years.
The methods rely upon measurements of entirely different 
quantities that are then applied using completely different
astrophysical arguments. 
Thus, any systematic effects are likely to be quite different
for the two types of sources. The fact that they yield quite
similar results is encouraging.  

\section{Summary}

A sample of thiry radio galaxies was used to determine 
cosmological parameters and the radio galaxy model 
parameters $\kappa_{RG}$ and $\beta$
in three standard cosmological models. Nearly
identical values of $\kappa_{RG}$ and $\beta$ are obtained 
in each cosmological model indicating that they are not
strongly affected by the context in which they are determined
(see Tables 1, 2, and 3), thus they can be used to determine
the dimensionless coordinate distance to each source.

Three supernovae samples are considered both separately and jointly
with the radio galaxy sample and are analyzed in the context
of three standard cosmological models (see Tables 1, 2, and 3)
to determine cosmological parameters and the model parameter
$\kappa_{SN}$.  Nearly identical values of $\kappa_{SN}$ are
obtained for each supernovae sample in the context of 
each cosmological model, thus they can be used to 
determine the dimensionless coordinate distance to 
each source.  They can also be used to determine the 
effective distance modulus
to each of the radio galaxies, which can then be combined
with those of the supernovae to increase the sample sizes,
particularly at high redshift. 

Constraints on cosmological parameters obtained with radio galaxies
are consistent with, though weaker than, those obtained with supernovae
alone.  There are no inconsistencies between results obtained with 
radio galaxies and supernovae.  For example, in the context of
a standard quintessence model, 
radio galaxies alone indicate
that the universe is accelerating today with about 90 \% confidence 
(see Figure 6). The consistency between results obtained with radio 
galaxies alone, supernovae alone, and the combined supernovae and
radio galaxy samples suggests that neither method is plagued by 
unknown systematic errors; both methods seem to be working well. 
The radio galaxy and supernovae methods are completely independent, 
based on a completely different physics and observations,
and provide independent measures of distances to sources at similar redshift.
The facts that the cosmological parameters obtained in specific models
are consistent, and that the coordinate distances to sources at similar 
redshift are consistent suggests that systematic errors are not
playing a major role in either method.

Since nearly identical values of $\kappa_{SN}$, $\kappa_{RG}$, and $\beta$
are obtained in each cosmological model, they can be used to solve for
the dimensionless coordinate distance to each source, $y$.  Values
of $y$ to each radio galaxy, obtained using the 
best fit values of $\kappa_{RG}$ 
and $\beta$ indicated by fits to the joint sample of 192 supernovae and 
30 radio galaxies, are listed in Table 4. The best fit 
values of $y$ listed in Table 4 are combined with the best fit
value of $\kappa_{SN}$ obtained for the Davis et al. (2007) sample
to obtain the distance modulus to each of the radio galaxies,
$\mu_D$, which are listed in Table 4 and which can be added to 
those already listed in Davis et al. (2007).  Similarly, the 
values of $y$ to the radio galaxies obtained using the best
fit values of $\kappa_{RG}$ and $\beta$ indicated by fits to the joint 
sample of 182 supernovae and 30 radio galaxies were obtained
(but are not listed), and were combined with the best fit value of
$\kappa_{SN}$ obtained for the Riess et al. (2007) sample to 
obtain values of $\mu_R$ for the radio galaxies that can be combined
with those already listed by Riess et al. (2007).  
Since nearly identical values of $\kappa_{SN}$, $\kappa_{RG}$,
and $\beta$ and their uncertainties are obtained in each cosmological model,
the average of the values obtained in the different cosmological models 
and their uncertainties were used. 

New constraints were obtained on the model parameter $\alpha$
in the rolling scalar field model of Peebles \& Ratra (1988). 
These constraints are rather strong and indicate that 
$\alpha$ is close to zero for reasonable values of $\Omega_m$.
Thus, a cosmological constant provides a good description of the
data, and there is no indication that the energy density of the
dark energy is changing with redshift.  

New constraints were obtained on the radio galaxy model parameter
$\beta$, suggesting values of $\beta$ close to 1.5.  
This is interpreted in the context of a standard magnetic
braking model of energy extraction from a rotating black hole
(e.g. Blandford 1990).  This is a very
special value of $\beta$ for which the braking magnetic field
strength depends only upon the spin angular momentum per unit
mass and the gravitational radius of the black hole.  This 
suggests that when the magnetic field strength reaches this 
maximum or limiting value,
the relativistic outflow is triggered. The fact that the  magnetic field 
strength does not depend explicitly on the black hole mass for
this special value of $\beta$ may explain why it is that this
paricular type of radio source is able to provide a modified
standard yardstick for cosmological studies. 

We have provided further evidence that radio galaxies can be used to 
determine coordinate distances to sources, and thus cosmological parameters
through a standard angular diameter test. 
Our comparative study shows that values of cosmological parameters obtained
with radio galaxies are consistent with those obtained with supernovae.
Supernovae alone and radio galaxies alone both indicate that the
universe is accelerating at the current epoch when these data
are analyzed in specific models such as a 
a quintessence
model in spatially flat universe, a
lambda model in 
universe that allows for non-zero space curvature, and a rolling scalar field model
in a spatially flat universe.  All of these models rely upon 
the equations of General Relativity, and the results obtained in
these models would not be correct if General Relativity is not
the correct theory of gravity. These data are analyzed in a 
model-independent way that does not rely upon General Relativity
by Daly et al. (2008), who show that the supernovae data alone
and the radio galaxy data alone indicate that the universe is
accelerating at the current epoch independent of whether 
General Relativity is the correct theory of gravity.
When expressed as coordinate distances, both radio galaxy and supernova
samples can be combined, and used in a joint cosmological analysis,
as it was done, e.g., by Daly \& Djorgovski (2003, 2004).  We use these
expanded and combined samples in a separate paper (Daly et al. 2008).

\acknowledgements
We would like to thank the observers for their tireless efforts in 
obtaining the data used for this study.  We would also like to thank
the referee for providing very helpful suggestions on this work. 
This work was supported in part by U. S. National Science
Foundation grants AST-0507465 (R.A.D.) and AST-0407448 (S.G.D.),
and the Ajax Foundation (S.G.D.).

\begin{deluxetable}{lllllll}
\tablewidth{0pt}
\tablecaption{Cosmological Results in a Quintessence Model\label{cosmo}}
\tablehead{
\colhead{Sample} & \colhead{$w$} & \colhead{$\Omega_m$} &  \colhead{$\beta$} & 
\colhead{$\kappa_{SN}$} & \colhead{$\kappa_{RG}$}&
\colhead{$\chi^2$/dof}}
\startdata   
30 RG&  $-0.87 {}^{+0.3}_{-1.1}$&0-0.25& $1.35 \pm 0.2$&&$9.101 \pm
0.035$&28.9/26\\
192SN+30RG&$ -1.08 {}^{+.28}_{-.39}$& $0.29 {}^{+0.08}_{-0.11}$&$1.52 \pm
0.15$&$43.296 \pm 0.015$&$9.035 \pm 0.035$
&224.4/217\\
192SN&$-1.14 {}^{+0.29}_{-0.4}$&$0.31 {}^{+0.8}_{-0.1}$&&$43.295 \pm
0.015$&&194.2/189\\
182SN+30RG& $-1.66 \pm 0.66$ &$0.45 {}^{+ 0.05}_{-0.08} $& 
$1.55 {}^{+.15}_{-0.1}$&$43.356 \pm 0.016$&$9.029 \pm 0.035$
&186.8/207\\
182SN&$-1.74 \pm 0.74$&$0.46 \pm 0.06$&&$43.354 \pm 0.016$&&155.7/179\\
115SN+30RG&
$-0.91 {}^{+0.33}_{-0.53}$&$0.22 {}^{+0.16}_{-0.22}$&$1.50 \pm 0.15$&$43.158 \pm
0.015$&$9.038 \pm 0.035$
&142.6/140\\
115SN&$-1.08 {}^{+0.44}_{-0.57}$&$0.29 {}^{+0.13}_{-0.26}$&&$43.154 \pm
0.015$&&112.5/112\\
\enddata
\label{QTable}
\end{deluxetable}

\begin{deluxetable}{lllllll}
\tablewidth{0pt}

\tablecaption{Cosmological Results in a Lambda Model with Space
Curvature\label{cosmo}}
\tablehead{
\colhead{Sample} & \colhead{$\Omega_{\Lambda}$} & \colhead{$\Omega_m$} & 
\colhead{$\beta$} & 
\colhead{$\kappa_{SN}$} & \colhead{$\kappa_{RG}$}&
\colhead{$\chi^2$/dof} }
\startdata   
30 RG&  $0.85 {}^{+0.4}_{-1.3}$&0-0.24& $1.35 \pm 0.2$&&$9.112 \pm
0.035$&28.9/26\\
192SN+30RG&$0.8 \pm 0.18$& $0.30 \pm 0.09$&$1.52 \pm 0.15$&$43.296 \pm
0.015$&$9.034 \pm 0.035$
&224.4/217\\
192SN&$0.85 {}^{+0.16}_{-0.19}$&$0.33 {}^{+0.08}_{-0.1}$&&$43.293 \pm
0.015$&&194.0/189\\ 
182SN+30RG& $0.91 {}^{+ 0.16}_{-0.2}$&$0.45 \pm 0.09$& 
$1.60 {}^{+.1}_{-0.15}$&$43.365 \pm 0.016$&$9.046 \pm 0.035$
&186.9/207\\
182SN&$0.96 {}^{+0.16}_{-0.19}$&$0.48 {}^{+0.08}_{-0.09}$&&$43.362 \pm
0.016$&&155.6/179\\
115SN+30RG&
$0.69 {}^{+ 0.25}_{-0.32}$&$0.22 {}^{+0.17}_{-0.22}$&$1.50 \pm 0.15$&$43.157 \pm 0.015$&$9.039
\pm 0.035$
&142.6/140\\
115SN&$0.81 {}^{+0.28}_{-0.32}$&$0.31 {}^{+0.19}_{-0.22}$&&$43.153 \pm
0.015$&&112.5/112\\

\enddata
\label{KTable}
\end{deluxetable}

\begin{deluxetable}{lllllll}
\tablewidth{0pt}
\tablecaption{Cosmological Results in a Rolling Scalar Field Model\label{cosmo}}
\tablehead{
\colhead{Sample} & \colhead{$\alpha$} & \colhead{$\Omega_m$} & 
\colhead{$\beta$} & 
\colhead{$\kappa_{SN}$} & \colhead{$\kappa_{RG}$}&
\colhead{$\chi^2$/dof}}
\startdata   
30 RG&  $0-6.2$&0-0.25& $1.35 {}^{+0.2}_{-0.1}$&&$9.12 \pm 0.035$&29.1/26\\
192SN+30RG&$ 0 {}^{+1.25}_{-0}$& $0.27 {}^{+0.03}_{-0.12}$&$1.50
{}^{+0.15}_{-0.1}$&$43.301 \pm 0.015$&$9.032 \pm 0.035$
&224.5/217\\
192SN&$0 {}^{+0.95}_{-0}$&$0.27 {}^{+0.03}_{-0.1}$&&$43.304 \pm
0.015$&&194.3/189\\
182SN+30RG& $0 {}^{+0.5}_{-0}$&$0.34 {}^{+0.04}_{-0.07}$& 
$1.55 {}^{+0.15}_{-0.1}$&$43.394 \pm 0.016$&$9.009 \pm 0.035$
&188.5/207\\
182SN&$0 {}^{+0.45}_{-0}$&$0.34 {}^{+0.04}_{-0.06}$&&$43.394 \pm
0.016$&&157.9/179\\
115SN+30RG&$ 0.35 {}^{+3.7}_{-0.35}$& $0.21 {}^{+0.08}_{-0.15}$&$1.50 \pm
0.15$&$43.158 \pm 0.015$&$9.039 \pm 0.035$
&142.6/140\\
115SN&$0 {}^{+3.85}_{-0}$&$0.26 {}^{+0.04}_{-0.21}$&&$43.156 \pm
0.015$&&112.6/112\\
\enddata
\label{RTable}
\end{deluxetable}

\begin{deluxetable}{llllllll}
\tablewidth{0pt}
\tablecaption{Distances and Distance Moduli to 30 Radio Galaxies}
\tablehead{
\colhead{Source} & \colhead{z} & \colhead{$y$} &  \colhead{$\sigma_y$} & 
\colhead{$\mu_D$} & \colhead{$\sigma_{\mu(D)}$}&
\colhead{$\mu_R$} & \colhead{$\sigma_{\mu(R)}$}  }
\startdata  	
3C 239	&1.790	&1.37	&0.34	&46.20	&	0.54	&	46.12	&0.52	\\	
3C 322	&1.681	&1.31	&0.34	&46.02	&	0.56	&	45.95	&0.54	\\	
3C 68.2	&1.575	&1.57	&0.49	&46.34	&	0.67	&	46.26	&0.65	\\	
3C 437	&1.480	&0.93	&0.27	&45.11	&	0.63	&	45.05	&0.61	\\	
3C469.1	&1.336	&1.14	&0.33	&45.42	&	0.62	&	45.41	&0.60	\\	
3C 324	&1.210	&1.02	&0.31	&45.06	&	0.66	&	45.07	&0.64	\\	
3C 194	&1.190	&1.01	&0.20	&45.01	&	0.44	&	45.07	&0.43	\\	
3C 267	&1.144	&0.71	&0.14	&44.22	&	0.42	&	44.22	&0.41	\\	
3C 356	&1.079	&0.87	&0.18	&44.59	&	0.46	&	44.57	&0.44	\\	
3C 280	&0.996	&0.65	&0.12	&43.85	&	0.41	&	43.89	&0.40	\\	
3C 268.1&0.974	&0.75	&0.14	&44.14	&	0.42	&	44.14	&0.40	\\	
3C 289	&0.967	&0.59	&0.11	&43.64	&	0.40	&	43.67	&0.39	\\	
3C 325	&0.860	&0.71	&0.13	&43.89	&	0.40	&	43.92	&0.39	\\	
3C6.1	&0.840	&0.74	&0.09	&43.97	&	0.25	&	43.99	&0.24	\\	
3C54	&0.827	&0.76	&0.08	&44.00	&	0.24	&	44.02	&0.23	\\	
3C114	&0.815	&0.64	&0.07	&43.61	&	0.24	&	43.62	&0.23	\\	
3C 265	&0.811	&0.59	&0.08	&43.44	&	0.29	&	43.48	&0.28	\\	
3C41	&0.794	&0.63	&0.07	&43.57	&	0.25	&	43.63	&0.24	\\	
3C 247	&0.749	&0.54	&0.07	&43.18	&	0.27	&	43.24	&0.27	\\	
3C 55	&0.720	&0.59	&0.08	&43.31	&	0.29	&	43.30	&0.28	\\	
3C441	&0.707	&0.53	&0.07	&43.10	&	0.27	&	43.12	&0.26	\\	
3C34	&0.690	&0.59	&0.06	&43.30	&	0.24	&	43.31	&0.23	\\	
3C44	&0.660	&0.76	&0.08	&43.81	&	0.24	&	43.80	&0.23	\\	
3C169.1	&0.633	&0.62	&0.07	&43.33	&	0.25	&	43.36	&0.24	\\	
3C 337	&0.630	&0.51	&0.07	&42.88	&	0.30	&	42.94	&0.29	\\	
3C 330	&0.549	&0.34	&0.07	&41.92	&	0.41	&	41.96	&0.40	\\	
3C172	&0.519	&0.66	&0.14	&43.31	&	0.45	&	43.30	&0.43	\\	
3C 244.1&0.430	&0.36	&0.07	&41.87	&	0.40	&	41.95	&0.39	\\	
3C142.1	&0.406	&0.33	&0.06	&41.65	&	0.40	&	41.69	&0.38	\\	
3C 405	&0.056	&0.05	&0.01	&36.97	&	0.44	&	37.02	&0.43	\\
\enddata
\tablecomments{Values of $\mu_D$ were obtained for the radio galaxies using
the best fit values of $\kappa_{SN}$, $\kappa_{RG}$, and $\beta$ obtained from 
the joint 192 supernovae and 30 radio 
galaxy fits, using the average of the values of these 
parameters listed in Tables 1, 2, and 3.
Similarly, 
values of $\mu_R$ were obtained using the best fit values
of $\kappa_{SN}$, $\kappa_{RG}$, and $\beta$ 
obtained from the 
joint fits to 182 supernovae and 30 radio galaxies, using the 
average of the values listed in Tables 1, 2, and 3. }
\label{yTable}
\end{deluxetable}

\clearpage

\clearpage

\clearpage

\clearpage

\clearpage

\clearpage

\clearpage

\clearpage

\clearpage

\end{document}